
\def\sqr#1#2{{\vcenter{\hrule height.#2pt
	     \hbox{\vrule width.#2pt height#1pt \kern#1pt
	     \vrule width.#2pt}
	     \hrule height.#2pt}}}
\def\boxx{\hskip1pt\mathchoice\sqr54\sqr54\sqr{3.2}4\sqr{2.6}4
					       \hskip1pt}
\font\eightrm=cmr8               
                
\font\eightbf=cmbx8
\def\pmb#1{\setbox0=\hbox{#1}\kern-.025em
    \copy0\kern-\wd0\kern.05em\kern-.025em\raise.029em\box0}

\def\sc#1{\hbox{\sc #1}}

\def\a{\alpha}      \def\l{\lambda}   \def\L{\Lambda}
\def\b{\beta}       \def\m{\mu}
\def\g{\gamma}      \def\G{\Gamma}    
\def\d{\delta}      \def\D{\Delta}    \def\p{\pi}     
                      \def\r{\rho}
\def\ve{\varepsilon}                  \def\s{\sigma}  
                         
\def\th{\theta}        \def\vphi{\varphi}
                   \def\o{\omega}   
\def\k{\kappa}

             \def\cH{{\cal H}}
             
\def\fr#1 #2{\hbox{${#1\over #2}$}}
\def\section#1{
  \vskip.7cm\goodbreak
  \centerline{\bf \uppercase{#1}}
  \nobreak\vskip.4cm\nobreak  }
\def\subsection#1{
  \vskip.7cm\goodbreak
  \noindent{\bf #1}
  \nobreak\vskip.4cm\nobreak  }

\def\sub#1{\vskip.3cm\goodbreak\noindent{\it #1}\vskip.3cm\nobreak}

\def\ftn{\eightrm\baselineskip=9pt}

\def\ver#1{\left\vert\vbox to #1mm{}\right.}


\magnification=\magstep1
\null

\section{Hamiltonian analysis of $SL(2,R)$ symmetry in Liouville
theory $^\ast$}
\footnote{}{$\ast\,$\ftn Lectures presented at the Danube Workshop
'93, June 1993, Belgrade, Yugoslavia.}

\vskip.5cm
\centerline{\it M. Blagojevi\' c}
\centerline{\it Institute of Physics, P.O.Box 57, 11001 Belgrade,
              Yugoslavia}

\centerline{\it and}

\centerline{\it T. Vuka\v sinac}
\centerline{\it Department of Theoretical Physics, Institute for
                Nuclear Sciences, Vin\v ca	}
\centerline{\it P.O.Box 522, 11001 Belgrade, Yugoslavia}
\vskip.5cm

\centerline{\eightbf Abstract}
\vskip.2cm
{\eightrm  We consider a Hamiltonian analysis of the Liouville
theory and construction of symmetry generators using Castellani's
method. We then discuss the light-cone gauge fixed theory.
In particular, we clarify the difference between Hamiltonian
approaches based on different choices of time,  $\xi^0$
and $\xi^+$. Our main result is the construction of
SL(2,R) symmetry generators in both cases. }

\subsection{1. Introduction} 

Einstein's theory of gravity in 2d is dynamically trivial, as the
corresponding action is a topological invariant. By studying the
quantum fluctuations of matter fields coupled to the 2d metric (treated
as an independent dynamical variable), Polyakov [1] obtained the
nontrivial effective theory of gravity in the form
$$
W[g] = \k \int d^2\xi \sqrt{-g}\biggl( {\fr 1 2}R{1\over\boxx}R
+\m^2 \biggr) \, .
$$
The original theory of matter couplings is invariant under general
coordinate transformations, as well as Weyl rescalings. If the quantum
fluctuations of matter fields are regulated so as to preserve general
coordinate invariance, the Weyl invariance is lost and the effective
action is related to the notion of the Weyl anomaly.

Polyakov and his collaborators [2] demonstrated that in the light-cone
gauge the $n$-point functions of the effective theory can be explicitly
found.  Although the gauge is fixed, these solutions display an
$SL(2,R)$ symmetry.  The origin of the symmetry, in the case where
the theory is also Weyl invariant, has been traced by Dass
and Summitra [3] to the residual symmetry transformations that leave
the light-cone gauge intact. They showed that after fixing the
light-cone gauge there exists a combination of general coordinate
transformations and Weyl transformations that is still a symmetry of
the theory --- the $SL(2,R)$ symmetry alluded to before.  The results
of the light-cone analysis inspired the investigation of the 2d gravity
in the conformal gauge (David, Distler and Kawai, [4]).  Interesting
Lagrangian treatment of the (light-cone and conformal) symmetris of the
theory was given in [5], while the BRST quantization was discussed in
[6].

Although many important properties of the 2d gravity have been obtained
in the path-integral quantization sheeme, it is also important to
understand the Hamiltonian structure of the theory. Egorian and
Manvelian [7] noted that the correct approach to understanding the
residual symmetry is to use the light-cone coordinate $\xi^+$ as the
time variable in the Hamiltonian approach. They showed that the
constraints appearing in the total Hamiltonian satisfy the $SL(2,R)$
algebra.  Although the $SL(2,R)$ symmetry is the symmetry of the
light-cone gauge, Abdalla et al. [8] were able to find a canonical
description of the theory in terms of gauge independent variables ---
$SL(2,R)$ currents, which is very important for understanding the
general structure of the theory, and, in particular, the relation to
the conformal gauge. The currents are defined in a gauge independent
way, but they are conserved only in the light-cone gauge. These
variables were used by Mikovi\' c [9] to derive some exact results in
2d gravity in a gauge independent way.  Ghosh and Mukherjee [10] used
the improved Hamitonian formalism in which the generators of
reparametrizations are corrected by certain surface terms,
and showed that these terms represent the $SL(2,R)$ currents.

The residual $SL(2,R)$ symmetry is not the standard local symmetry: it
is characterized by parameters $\o_a$ which are not arbitrary
functions of both coordinates, but depend only on $\xi^+$.  The fact
that this symmetry appears in the gauge-fixed theory seems to contradict
the basic principle of the Dirac formalism [11], since there  is a
symmetry defined by parameters $\o_a(\xi^+)$, but there are no
first-class constraints, as the gauge is fixed. Barcelos-Neto [12]
noted that the contradiction is only an apparent one, but his arguments
were not complete.

The subject of the present paper is a complete analysis of the
$SL(2,R)$ symmetry  of the effective 2d gravity as a (nonstandard)
Hamiltonian local symmetry, including the construction of the
corresponding generators as well as the comparison with the Noether
method and the improved Hamiltonian approach. In particular, we clarify
the difference between the Hamiltonian aproaches based on times $\xi^0$
and $\xi^+$, respectively. In Sec. 2 we develop the general Hamiltonian
description of the covariant theory and describe the light-cone gauge.
In Sec. 3 we discuss the $SL(2,R)$ symmetry of the gauge fixed theory
in the light-front
formalism with $\xi^+$ as the time variable, and construct the related
gauge generators by using Castellani's algorithm [13]. In
Sec. 4  previous results are translated into the standard formalism
based on the time $\xi^0$.  Section 5 is devoted to conclusions. Some
technical details and a modification of Castellani's method appropriate
to the case of residual symmetries are discussed in the Appendix.

\subsection{2. The standard Hamiltonian analysis} 

\sub{A. Constraints and the Hamiltonian} 

The Liouville action $I=-W/\kappa$ can be written as a local functional
by introducing an auxilliary field $F$ [14] :
$$
I[F,g]= -\int d^2\xi\sqrt{-g}\bigl[
 {\fr 1 2}g^{\a\b}\partial_\a F\partial_\b F
 +{\a\over 2}FR + \mu^2 \bigr] \, .                           \eqno(1a)
$$
The elimination of $F$ with the help of the equations of motion leads
back to the nonlocal theory.  After disregarding a surface term the
Liouville action takes the following form [7,12]:
$$
\eqalign{
I[F,g_{\a\b}]=&\int d^2\xi{1\over\sqrt{-g}}
 \biggl\{ {\fr 1 2}(g_{11}{\dot F}^2+g_{00} F^{\prime 2})
 -g_{01}{\dot F} F^\prime  \cr
&+{\a\over 2}\biggl[ {g_{01}\over g_{11}}({\dot F}
 g_{11}' - F '{\dot g}_{11}) + {\dot g}_{11}{\dot F} +g_{00}' F '
 -2g_{01} '{\dot F}\biggr] + g\mu^2 \biggr\} \, .}            \eqno(1b)
$$

The basic Lagrangian dynamical variables are $ (F,g_{\a\b})$.
The corresponding canonical momenta $(\pi_F,\pi^{\a\b})$ are
easily obtained in the form
$$\eqalign{
&\pi_ F ={1\over\sqrt{-g}}\biggl[ g_{11}{\dot F} - g_{01} F'
 + {\a\over 2}\biggl( {g_{01}\over g_{11}}g_{11}'
 +{\dot g}_{11}-2g_{01}'\biggr)\biggr] \, ,\cr
&\pi^{11}= {\a\over 2\sqrt{-g}}
 \biggl( {\dot F}-{g_{01}\over g_{11}} F' \biggr) \, , \cr
&\pi^{01}=0 \, ,\qquad\qquad \pi^{00}=0 \, ,\cr}              \eqno(2)
$$
leading to the following primary constraints:
$$
\phi^{01}\equiv\pi^{01}\approx 0 \, ,\qquad
\phi^{00}\equiv\pi^{00}\approx 0 \, .
$$

Having found all the primary constraints we now proceed to find the
Hamiltonian. It is convenient to write the Lagrangian in a condensed
form as
$$
{\cal L} = a{\dot F}^2 + b{\dot F}{\dot g}_{11}
	     +c{\dot F} + d{\dot g}_{11} + e \, .
$$
The corresponding canonical Hamiltonian density is
$$
{\cal H}_c=\pi_F{\dot F}+\pi^{11}{\dot g}_{11}-{\cal L}
= -{a\over b^2}(\pi^{11}-d)^2 + {1\over b}(\pi^{11}-d)(\pi_F -c)-e\, ,
$$
or, more explicitly,
$$
{\cal H}_c= -{\sqrt{-g}\over g_{11}}{\cal H}_0
	    + {g_{01}\over g_{11}}{\cal H}_1 \, ,             \eqno(3a)
$$
where
$$\eqalign{
&{\cal H}_0={1\over 2}\biggl[ {4\over\a^2} (g_{11}\pi^{11})^2
 -{4\over\a}g_{11}\pi^{11}\pi_F - F'^2 + 2\a F''
 -\a{g'_{11}\over g_{11}} F'-2g_{11}\mu^2\biggr]\, ,\cr
&{\cal H}_1=\pi_ F F'-2g_{11}(\pi^{11})'-\pi^{11}g_{11}'\, .} \eqno(3b)
$$
The general Hamiltonian dynamics of the system is described by the total
Hamiltonian density,
$$
{\cal H}_T={\cal H}_c + u_{01}\phi^{01} + u_{00}\phi^{00} \, ,\eqno(4)
$$
where $u_{0 1}$ and $u_{0 0}$ are, at this stage, arbitrary multipliers.

In order to have a consistent Hamiltonian theory we shall demand that
the constraints do not change during the general time evolution of the
dynamical system governed by the total Hamiltonian $H_T\equiv\int
d^2\xi{\cal H}_T$. Using the usual Poisson brackets beetwen the basic
Hamiltonian variables, the consistency of the primary constraints leads
to
$$
\eqalign{
&{\dot \phi}^{01}=\bigl\{ \phi^{01}, H_T\bigr\} = {1\over g_{11}}
  \biggl( -{\cal H}_1+{g_{01}\over\sqrt{-g}} {\cal H}_0\biggr) \, ,\cr
&{\dot \phi}^{0 0}=\bigl\{ \phi^{0 0}, H_T\bigr\} =
  -{1\over 2\sqrt{-g}} {\cal H}_0 \, .}
$$
Therefore, the secondary constraints are just ${\cal H}_0$ and
${\cal H}_1$.

A direct calculation yields the following Poisson brackets between
${\cal H}_0$ and ${\cal H}_1$:
$$\eqalign{
&\{ {\cal H}_0 (\sigma ), {\cal H}_0 (\sigma ')\} =
[{\cal H}_1(\sigma )+{\cal H}_1 (\sigma ')]\partial_\sigma\delta\, ,\cr
&\{ {\cal H}_0 (\sigma ), {\cal H}_1 (\sigma ')\} =
[{\cal H}_0(\sigma )+{\cal H}_0(\sigma ')]\partial_\sigma\delta \, ,\cr
&\{ {\cal H}_1 (\sigma ), {\cal H}_1 (\sigma ')\} =
[{\cal H}_1(\sigma )+{\cal H}_1 (\sigma ')]\partial_\sigma\delta \, , }
							      \eqno(5)
$$
where $\d\equiv \d (\s ,\s ')$.  It is now easy to see that the
consistency of ${\cal H}_0$ and ${\cal H}_1$ does not produce further
constraints.

\sub{B. The reparametrization symmetry} 

All the constraints $\phi^{00}, \phi^{01}, {\cal H}_0$ and ${\cal H}_1$
are of the {\it first class}, which implies the existence of a {\it
gauge symmetry}.  One can study more explicitly the nature of the
symmetry by constructing the corresponding gauge generators, and
calculating the related gauge transformations of dynamical variables.

A general method of constructing the generators of a local symmetry in
the Hamiltonian approach has been given by Castellani [13]. Here we
shall limit ourselves to physically important situations where gauge
transformations are given in terms of arbitrary parameters $\ve$ and
their first time derivatives $\dot\ve$. In that case the gauge
generators take the form
$$
G[\ve ]= \int d\xi^1\bigl(\ve G_0 + \dot\ve G_1 \bigr) \, ,   \eqno(6a)
$$
where $G_0$ and $G_1$ are phase-space functions satisfying the
conditions
$$\eqalign{
G_1 &=C_{PFC} \, ,\cr
G_0 + \{ G_1,H_T\} &=C_{PFC} \, ,\cr
\{ G_0,H_T\} &=C_{PFC} \, ,}                                  \eqno(6b)
$$
and $C_{PFC}$ means primary first-class (PFC) constraint.

Starting with the PFC constraints $2g_{00}\p^{00}+g_{01}\p^{01}$ and
$g_{11}\p^{01}+2g_{01}\p^{00}$ as $G_1$'s, Castellani's method
yields the following expressions for the gauge generators:
$$\eqalign{
&G[\ve^0]= \int d\xi^1 \bigl\{ \ve^0 [\cH_T -(g_{00}\p^{01})']
	  + {\dot\ve}{^0}(2g_{00}\p^{00}+g_{01}\p^{01}) \bigr\}\, ,\cr
&G[\ve^1]= \int d\xi^1 \bigl\{ \ve^1 [{\cal H}_1 +g_{00}'\p^{00}
						    -g_{01}(\p^{01})']
   +{\dot\ve}{^1}(g_{11}\p^{01}+2g_{01}\p^{00}) \bigr\} \, ,} \eqno(7)
$$
where $\ve^\a=\ve^\a (\xi^0,\xi^1)$. The corresponding gauge
transformations are defined by
$$
\d_0 X =\{ X , G\}\, ,\qquad G\equiv G[\ve^0]+G[\ve^1]\, .
$$
Their form on the fields $(F,g_{\a\b})$ is easily seen to coincide
with the reparametrizations:
$$\eqalign{
&\d_0 g_{\a\b}={\ve^\g}_{,\a}g_{\g\b} +
{\ve^\g}_{,\b}g_{\g\a}+\ve^\g g_{\a\b ,\g} \, ,\cr
&\d_0 F=\ve^\a\partial_\a F \, .}
$$

\sub{C. The light-cone gauge} 

The theory (1a) simplifies significantly in the light-cone gauge,
defined by
$$\eqalign{
&g_{+-}=1 \, ,\qquad g_{--}= 0 \, , \cr
&ds^2=2h(d\xi^+)^2+2d\xi^+d\xi^-\, , }                        \eqno(8)
$$
where $g_{ab}$ $(a,b=+,-)$ are the components of the metric tensor in
the light-cone coordinates $\xi^\pm=(\xi^0\pm \xi^1)/\sqrt{2}$ :
$$\eqalign{
&g_{+-}\equiv {\fr 1 2}(g_{00}-g_{11})\, ,\qquad
g_{--}\equiv {\fr 1 2}(g_{00}+g_{11})-g_{01} \, ,\cr
&2h\equiv g_{++}={\fr 1 2}(g_{00}+g_{11})+g_{01} \, .}
$$
After finding the inverse metric
$$
g^{++}=0\, ,\qquad g^{+-}=g^{-+}=1\, ,\qquad  g^{--}=-2h \, ,
$$
the calculation of the metric connection $\G^\a_{\b\g}$
yields the following nonvanishing components:
$$ \eqalign{
&\G_{++}^+=-\partial_-h \, ,\qquad \G_{+-}^-=\partial_-h \, ,\cr
&\G_{++}^-=\partial_+h + 2h\partial_-h\, ,}
$$
where $\partial_\pm =(\partial_0\pm\partial_1)/\sqrt{2}$.
The curvature components and the laplacian are of the simple form
$$\eqalign{
&R_{++}=2h\partial_-^2h\, ,\qquad R_{+-}=\partial_-^2h\, ,
\qquad R=2\partial_-^2h\, ,\cr
&{\boxx}=2\partial_-(\partial_+-h\partial_-)\, . }
$$
The equation of motion for $F$ is given by
$$
{\boxx} F = {\a\over 2}R \qquad \Longrightarrow \qquad
\partial_-(\partial_+-h\partial_-)F = {\a\over 2}\partial_-^2h \, .
$$
The equations of motion for $g_{\a\b}$ can be obtained from the
energy-momentum corresponding to the action (1a). By using the relation
$$
\d \int d^2\xi\sqrt{-g}FR=-\int d^2\xi\sqrt{-g}\d g^{\a\b}
			  (\nabla_\a\nabla_\b -g_{\a\b}\nabla^2 )F \, ,
$$
one easily finds
$$
T_{\a\b}=-{\fr 1 2}\partial_\a F\partial_\b F
  +{\a\over 2}\bigl(\nabla_\a\nabla_\b -g_{\a\b}\nabla^2 \bigr) F
  +{\fr 1 2}g_{\a\b} \bigl({\fr 1 2}g^{\g\d}\partial_\g F\partial_\d F
  +\m^2 \bigr) \, .
$$
In the light-cone gauge they are reduced to
$$\eqalign{
& T_{--} = -{\fr 1 2}(\partial_-F)^2 +{\a\over 2}\partial_-^2F  \, ,\cr
& T_{+-} = hT_{--}-{\fr 1 4}(\a^2\partial_-^2h-2\mu^2)  \, ,\cr
& T_{++} = h^2 T_{--} -h(\a^2\partial_-^2h-2\mu^2) +{\a^2\over 8}
 \bigl[ (\partial_-h)^2 -2h\partial_-^2h+2\partial_-\partial_+h\bigr]
						    \, ,}     \eqno(10)
$$
after using the equation of motion for $F$. From the Eq.(9) and $T_{--}=0$
follows the important result
$$
\partial_-^3h=0\, .                                           \eqno(11)
$$
Now we shall focus our attention to {\it the gauged-fixed action}, which has
the following form:
$$
 I=\int d^2\xi\bigl[-\partial_+F\partial_-F +
  h(\partial_-F)^2-\a F\partial_-^2h -\mu^2\bigr] \, .        \eqno(12)
$$
As a result we have only two equations of motion: Eq.(9) and $T_{--}=0$
and we shall show that they are $SL(2,R)$ invariant, which is not the case
for the whole set of equations of motion [5].

For the begining, let us observe that the gauge-fixed action (12) is
{\it nondegenerate} :
$$\eqalign{
&\p_h={\a\over 2} (\dot F-F')\, ,\cr
&\p=-\dot F +h(\dot F-F')+{\a\over 2}(\dot h-h')\, .}
$$
This can be also seen from the relations (2) restricted to the
light-cone gauge:
$$\eqalign{
&\p^{11}+{\a\over 2}{F'\over g_{11}}={\a\over 2}(\dot F-F')\, ,\cr
&\p_F-{\a\over 2}{g_{11}'\over g_{11}}=-\dot F +h(\dot F-F')
				      +{\a\over 2}(\dot h-h')\, .}
$$
These two equations show that the transition
$(\p^{11},\p_F)\to (\p_h,\p )$ is achived by a canonical
transformation
$$
\p_h = \p^{11}+{\a\over 2}{F'\over g_{11}}\, ,\qquad
\p = \p_F-{\a\over 2}{g_{11}'\over g_{11}} \, .
$$
The reason for this lies in the fact that the action
(1b) is obtained after disregarding a surface terms in (1a), whereas
Eq.(12) is obtained directly from (1a).

The nondegeneracy of the action (12) is a natural consequence of the
gauge fixing procedure, and implies the absence of first class
constraints. Consequently, we can conclude
that there are no gauge symmetries of the Hamiltonian equations of motion.

The true meaning of this assertion is the following. It is well known
that gauge symmetries in the Hamiltonian framework are related to the
presence of arbitrary multipliers in the total Hamiltonian. Let us
consider a dynamical evolution of a system described by a phase-space
trajectory starting from a given point at time $t=0$. For different
choices of arbitrary multipliers we can solve the Hamiltonian equations
of motion and obtain different trajectories, all starting from the same
point and describing the same physical state. At any time $t>0$ we can
pass from one trajectory to the other, without changing the physical
state. This unphysical transition from one trajectory to the other at a
given time $t$ is called the gauge transformation. It is clear that the
Hamiltonian definition of gauge symmetries is based on a {\it definite
choice of time\/.} The absence of gauge symmetries in a given
Hamiltonian formalism based on one specific choice of time does not
mean that these symmetries are also absent for any other choice. We
shall see in the next section how the hidden symmetries of the
Liouville theory are detected by using the light-cone time variable
$\xi^+$.

\subsection{3. SL(2,R) symmetry in the light-front formalism} 

There are several reasons to study relativistic field theories at fixed
light-cone time $\xi^+$. Dirac [15] showed that in the light-front form
of the Hamiltonian formalism a maximum number of Poincare generators
becomes independent of the dynamics. The same approach was used to
develop a practical method of performing non-perturbative calculations
in quantum field theory, and to study the problem of vacuum structure
[16]. Here, the light-front formalism is used to clarify the nature of
the residual symmetries in the Liouville theory.

\sub{A. Hamiltonian analysis} 

We have seen in the standard Hamiltonian approach that the Liouville
action in the light-cone gauge is not degenerate. However, if we choose
$\xi^{+}$ as the time variable, then the action (12) becomes degenerate.
The definition of the momenta $(\p,\p_h)$ corresponding to the
Lagrangian variables $(F,h)$ leads to the following primary
constraints:
$$
\vphi_1\equiv\pi_h\approx 0 \ ,\qquad \vphi_2\equiv\pi +\partial_{-}F
				      \approx 0  \, .         \eqno(13)
$$
The canonical Hamiltonian density is
$$
{\cal H}_c= -h(\partial_-F)^2-\a\partial_-F\partial_-h+\mu^2 \, ,
$$
while the total Hamiltonian density takes the form
$$
{\cal H}_T= {\cal H}_c + u_1\vphi_1 + u_2\vphi_2 \, .
$$

The consistency requirements are calculated by using the Poisson
brackets taken at the same time $\xi^+$, and lead to further
constraint. By demanding $\{\vphi_1,H_T\} =0$ one obtaines the
secondary constraint
$$
\chi_1\equiv (\partial_-F)^2-\a\partial^2_-F \approx 0 \, ,
$$
the consistency of $\chi_1$ yields the tertiary constraint
$$
\th_1\equiv -{\a^2\over 2}\partial^3_-h \approx 0  \, ,
$$
while the consistency of $\th_1$ leads to the condition on $u_1$:
$$
\partial^3_-u_1=0 \, .
$$
The last relation can be solved in the form
$$
u_1(\xi^+,\xi^-)= u_-(\xi^+)+\xi^-u_0(\xi^+)+(\xi^-)^2u_+(\xi^+) \, ,
$$
where $u_-,u_0,u_+$ are arbitrary functions of the time $\xi^+$.

On the other hand, the requirement $\{\vphi_2,H_T\} =0$ leads to
$$
u_2={\hat u}_2+v(\xi^+)\, ,\qquad
{\hat u}_2 \equiv h\partial_-F+{\a\over 2}\partial_-h  \, ,
$$
where $v$ is an arbitrary multiplier depending on $\xi^+$ only.

Now, the total Hamiltonian can be written in the form
$$
H_T= H' +u_-\vphi^- + u_0\vphi^0 + u_+\vphi^+ +v\rho \, ,    \eqno(14)
$$
where
$$
H'=\int d^2\xi^-\bigl( {\cal H}_c +{\hat u}_2\vphi_2 \bigr) =
    \int d\xi^-\, \bigl[ -{\a\over 2}\partial_-h\partial_-F +
    {\hat u}_2 \pi +\mu^2\bigr] \, ,                         \eqno(15a)
$$
and
$$\eqalign{
&\vphi^a\equiv\int d\xi^-\, (\xi^-)^{a+1}\vphi_1
	  = \int d\xi^-\, (\xi^-)^{a+1}\pi_h   \, ,  \cr
&\rho\equiv\int d\xi^-\, \vphi_2 = \int d\xi^-\,
				 (\pi +\partial_-F ) \, ,}   \eqno(15b)
$$
with $a=(-1,0,+1)$. The algebra of the constraints is
$$\eqalign{
\{\vphi_1(\xi_1^-),\th_1(\xi_2^-)\}&={\a^2\over 2}\partial_-^3\d\, ,\cr
\{\vphi_2(\xi_1^-),\chi_1(\xi_2^-)\}&= -2\partial_-F\partial_-\d
			      +\a\partial^2_-\delta \, ,\cr
\{\vphi_2(\xi_1^-),\vphi_2(\xi_2^-)\}&=2\partial_-\d  \, ,}    \eqno(16)
$$
where $\delta\equiv\delta (\xi_1^- -\xi_2^-)$, and all other Poisson
brackets vanish. Although these constraints are not of the first class
their combinations, given by Eq.(15b), might be, which can be seen from
Eq.(14). Since we are now dealing with nonlocal quantities we have to
check whether they have well defined functional derivatives.

\sub{B. Asymptotic behaviour and surface terms} 

The solution of the problem is given by the following considerations
[17,11]. In field theory the Hamiltonian and the gauge generators
are expressed as functionals of the phase-space variables,
$$
G[q,p]=\int dx\,{\cal G}[q(x),\partial_\a q(x),
					   p(x),\partial_\a p(x)]\, .
$$
Since $G[q,p]$ is a {\it nonlocal} expression that acts on phase-space
variables via the Poisson brackets, one has to check whether this
quantity has well defined {\it functional derivatives\/.}

The first step in that direction is to define precisely the phase space
in which all the nonlocal quantities act. This is achieved by defining
the {\it asymptotic behaviour} of the basic dynamical variables.

The constraints $\chi_1$ and $\th_1$  can be easily solved leading to
$$\eqalign{
&F(\xi^+,\xi^-)=-(\a /2)\ln\partial_-f \, ,
		\qquad f\equiv (a\xi^-+b)/(c\xi^-+d) \, ,\cr
&h(\xi^+,\xi^-)=J^+ -2\xi^-J^0 + (\xi^-)^2J^-   \, , }       \eqno(17a)
$$
where $a,b,c,d$ and $J^a$ are arbitrary functions of $\xi^+$, and
$ad-bc=1$. From
these solutions we find the following assymptotic behaviour of the
field variables $(F,h)$:
$$\eqalign{
&F = -\a\ln \vert\xi^-\vert + A(\xi^+) + O_1 \, , \cr
&h = J^+(\xi^+)-2\xi^-J^0(\xi^+)+(\xi^-)^2J^-(\xi^+)\, , }   \eqno(17b)
$$
where $O_n$ denotes a term that decreases like $(\xi^-)^{-n}$ or faster
for large $\xi^-$, i.e. $(\xi^-)^nO_n$ remains finite when  $\xi^-\to
\infty$.

It should be noted that for those expressions that vanish on shell one
can demand an arbitraryly fast decrease, as no solution of the
equations of motion is thereby lost. In accordance with this remark the
asymptotic behaviour of the momentum variables is determined by
requiring
$$
p-{\partial{\cal L}\over \partial\dot q}=\hat O \, ,
$$
where $\hat O$ denotes a term that decreases sufficiently fast, e.g.
like $O_3$. By using the definitions of momenta (13) and the accepted
asymptotic behaviour of the fields, one finds
$$
\p_h = \hat O \, ,\qquad \p = {\a\over \xi^-} + O_2 \, .     \eqno(18)
$$

Keeping in mind the above asymptotic relations we now wish to
check whether various nonlocal expressions in the theory have
functional derivatives. A functional $G[q,p]$ has well defined
functional derivatives if its variation can be written as
$$
\d_0G=\int dx \bigl[ A(x)\d_0 q(x) + B(x)\d_0 p(x) \bigr]\, ,\eqno(19)
$$
where $\d_0 q_{,\a}$ and $\d_0 p_{,\a}$ are absent. In general,
when the derivatives of fields are present in $G$, this
requirement will not be satisfied. This will lead us to redefine $G$
by adding certain surface terms, obtaining in this way correctly
defined quantities. If the surface terms happen to vanish, then the
original functional does not need any modification.

Let us proceed by demonstrating the above procedure in the case of the
canonical Hamiltonian. The variation of $H_c$ yields
$$\eqalign{
\d_0 H_c &= \int d\xi^-\bigl[ -2h\partial_-f(\partial_-\d_0 F)-\a
(\partial_-\d_0 h)\partial_- F -\a \partial_-h(\partial_-\d_0 F)\bigr]
							       + R \cr
&=-2h\partial_-F\d_0 F -\a\d_0 h\partial_-F-\a\partial_-h\d_0 F
				\ver{3}^{+\infty}_{-\infty} + R \, ,}
$$
where those terms that contain the unwanted variations of fields or
momenta are explicitly displayed, and the remaining terms of the
correct form (19) are denoted by $R$. The second equality is obtained
by performig the integration by  parts, and this brings in the surface
terms. Denoting these terms by $S$ and using the asymptotics (17b) we
have
$$\eqalign{
S=&\bigl[ -2(J^+-2\xi^-J^0+(\xi^-)^2J^-)(-{\a\over\xi^-}+O_2)
  -\a (-2J^0+2\xi^-J^-)\bigr]\d_0 F \ver{3}^{+\infty}_{-\infty}   \cr
  &-\a (\d_0J^+-2\xi^-\d_0 J^0+(\xi^-)^2\d_0 J^-)(-{\a\over\xi^-}+O_2)
      \ver{3}^{+\infty}_{-\infty} \, .}
$$
{}From the asymptotic behaviour of $F$ and $h$ it follows
$$\eqalign{
&\d_0 F(+\infty )=\d_0 F(-\infty ) \, ,\cr
&\d_0 J^a(+\infty )=\d_0 J^a(-\infty ) \qquad (a=0,+,-) \, ,}
$$
and we easily obtain
$$
S=\a^2\xi^-\d_0 J^- \ver{3}^{+\infty}_{-\infty} \, .
$$
It is also easy to see that $H_c$ is not even finite but that it can be made
finite with well defined functional derivatives by adding a surface term
$$
{\tilde H}_c=H_c-\int d\xi^-({\a^2\over 2}\partial_-^2h+\mu^2)=
	    \int d\xi^-\bigl[ -h(\partial_-F)^2-\a\partial_-F\partial_-h
	    -{\a^2\over 2}\partial_-^2h\bigr]  \, .
$$
Similar considerations apply to $\r$ and $\vphi^a$, with the conclusion
that the whole total Hamiltonian ${\tilde H}_T$, where $H_c$ is replaced
with ${\tilde H_c}$, has correctly defined functional
derivatives, as all the surface terms vanish. Now, one can verify that
constraints $\r$ and $\vphi^a$, given by Eq.(15b), are {\it first-class}
since they have vanishing Poisson brackets with all "linear combinations"
of constraints
$$
\vphi [\l ]\equiv \int d\xi^-\l (\xi^+,\xi^-)\varphi (\xi^+,\xi^-)
$$
which are well defined (this requirement gives certain conditions on
parameters $\l$).

\sub{C. Construction of gauge generators} 

It is clear from Eq.(14) that the arbitrary multipliers in $H_T$ are
functions of the time $\xi^+$ only, in contrast to the general case
where they depend on both $\xi^-$ and $\xi^+$. The gauge generators
will be of the general form (6a), but the parameters must be of the
same type as the multipliers in $H_T$, $\o =\o (\xi^+)$:
$$
G[\o ]=\o (\xi^+) G_0 + \partial_+\o (\xi^+) G_1  \, .       \eqno(20a)
$$
A detailed analysis of Castellani's conditions shows that they might be
slightly changed in this case. Instead of (6b) we found that $G_0$ and
$G_1$ should satisfy the relations
$$\eqalign{
G_1 &={\tilde C}_{PFC} \, ,\cr
G_0 + \{ G_1,H_T\} &={\tilde C}_{PFC} \, ,\cr
\{ G_0,H_T\} &={\tilde C}_{PFC} \, ,}                        \eqno(20b)
$$
where ${\tilde C}_{PFC}$ denotes a {\it quantity multiplying an
arbitrary multiplier in} $H_T$. We see that in principle
it may happen that ${\tilde C}_{PFC}$ is not even a constraint. In our
case surface terms are absent, so ${\tilde C}_{PFC}$ {\it is} a
constraint.

Let us start with $G_1 = \vphi^a$ or $\r$ in (20b) and try to find the
corresponding $G_0$'s. The calculation of $\chi^a\equiv
\{\vphi^a,H_T\}$ leads to

$$
\chi^a =\int d\xi^-\, (\xi^-)^{a+1}\bigl[ -\pi\partial_-F
 -{\a\over 2}\partial_-(\partial_-F-\pi )\bigr]\approx 0 \, ,\eqno(21)
$$
while $\{\rho ,H_T\}=0$.

Before continuing, let us check on the differentiability of $\chi^a$.
By varying $\chi^a$ one finds
$$\eqalign{
\d_0\chi^a =& \int d\xi^-\, (\xi^-)^{a+1}\bigl[ -\pi\partial_-\d_0 F
 -(\a /2)\partial_-^2\d_0 F + (\a /2)\partial_-\d_0\pi )\bigr] +R \cr
=& (\xi^-)^{a+1}\bigl[ -\pi\d_0 F -(\a /2)\partial_-\d_0 F
   + (\a /2)\d_0\pi ) \bigr] +(\a /2)\partial_-(\xi^-)^{a+1}\d_0 F
   \ver{3}^{+\infty}_{-\infty} + R \, . }
$$
It is now easy to see that for $a=-1,0,1$ the above surface term
vanihes and , therefore, $\chi^a$ is differentiable.

The algebra of the constraints has the form
$$\eqalign{
\{ \chi^-,\chi^0\} &=-\chi^-    \, ,\cr
\{ \chi^-,\chi^+\} &=-2\chi^0   \, ,\cr
\{ \chi^0,\chi^+\} &=-\chi^+  \, ,}                          \eqno(22a)
$$
while the remaining Poisson brackets vanish. One also finds
$$\eqalign{
\{ \chi^-,H_T\} &= -2J^0\chi^- + 2J^-\chi^0 \, ,\cr
\{ \chi^0,H_T\} &= -J^+\chi^- + J^-\chi^+   \, ,\cr
\{ \chi^+,H_T\} &= -2J^+\chi^0 + 2J^0\chi^+ \, .}            \eqno(22b)
$$
The above relations represent the proof that  $\vphi^a$ and $\rho$ are
effectively PFC.

It should be noted that the constraints $\chi^a$ satisfy the $Sl(2,R)$
algebra, which is closely related to the residual symmetry
of the theory, as we shall see soon.

Now one can find $G_0$.  Starting with $\vphi^a$ as $G_1^a$,
one obtains
$$\eqalign{
G^-_0 &=-\chi^- -2J^0\vphi^- +2J^-\vphi^0 \, ,    \cr
G^0_0 &=-\chi^0 -J^+\vphi^- +J^-\vphi^+   \, ,    \cr
G^+_0 &=-\chi^+ -2J^+\vphi^0 +2J^0\vphi^+ \, ,  }            \eqno(23a)
$$
or, equivalently,
$$
G_0^a = -\chi^a+f^{abc}J_b\vphi_c   \, ,                     \eqno(23b)
$$
where $f^{abc}$ are the structure constants of $SL(2,R)$
($f^{abc}$ is totally antisymmetric, and $f^{+-0}=1$).

The complete gauge generator $G=G[\o_-]+G[\o_0]+G[\o_+]$ has the form
$$
G=\int d\xi^-\, \bigl[ (\partial_+\ve + \ve\partial_-h -h\partial_-\ve )
    \pi_h +(\ve\partial_-F+{\a\over 2}\partial_-\ve )\pi +
   {\a\over 2}(\partial^2_- \ve )F\bigr]  \, ,               \eqno(24)
$$
where we introduced the parameter
$$
\ve (\xi^+,\xi^-)=\o_-(\xi^+)+\xi^-\o_0(\xi^+)+(\xi^-)^2\o_+(\xi^+)\, ,
							     \eqno(25a)
$$
satisfying the relation
$$
\partial_-^3\ve =0\ \, .                                     \eqno(25b)
$$

The gauge transformations of the fields take the form
$$\eqalign{
\d_0 h\equiv\{ h,G\} &=\partial_+\ve
			 +\ve\partial_-h -h\partial_-\ve\, , \cr
\d_0 F\equiv\{ F,G\} &=\ve\partial_-F +{\a\over 2}\partial_-\ve \, , }
							     \eqno(26)
$$
which is easily recognized as the $SL(2,R)$ symmetry.

We note that the gauge generator obtained from $\rho$ has the form
$$
 G = \l (\xi^+)\rho(\xi^+)   \, ,
$$
and produces the trivial transformations of the fields:
$$
\d_0 h=0 \, ,\qquad \d_0 F=\l (\xi^+) \, .
$$

In this way, the $SL(2,R)$ symmetry of the Liouville theory in the
light-cone gauge is consistently described as a kind of gauge symmetry
in the light-front Hamiltonian formalism, based on  the time $\xi^+$.
The symmetry is described by three parameters $\o_a(\xi^+)$, which are
functions of only one coordinate --- the time $\xi^+$. The situation
differs from the case of standard gauge symmetries, where the gauge
parameters depend on both variables $\xi^+$ and $\xi^-$.  This property
is closely related to the the fact that the symmetry is a residual
symmetry of the theory.

\subsection{4. SL(2,R) symmetry in the standard formalism} 

Now we shall return to the time $\xi^0$ and try to understand the
existence of the $SL(2,R)$ symmetry in the Hamiltonian formalism based
on this "real" time. Although the gauge-fixed action is invariant under
the $SL(2,R)$ transformations, these transformations are {\it not the
gauge symmetries in the sense of the Hamiltonian formalism based on}
$\xi^0$, as we already explained at the end of Sec. 2 (the existence of
a gauge symmetry would contradict the fact that the Liouville theory in
the light-cone gauge is not degenerate).  In this situation it is
instructive to use a modification of Castellani's method and obtain
certain conditions that the generators of a global or residual (=not
local) symmetry of the Hamiltonian equations of motion should satisfy.
Unlike the case of local symmetries, these conditions do not give a
prescription for the construction of the generators. We shall also
analyse the problem from the point of view of the Noether currents.

The gauge fixed action (9) is invariant under the residual $SL(2,R)$
transformations (23) followed by the coordinate transformations
$\delta\xi^\mu =-\ve^\mu$, where $\ve\equiv\ve^-$. The total variation
of the Lagrangian density is given by
$$
\D {\cal L}\equiv \d {\cal L}-\partial_\mu ({\cal L}\ve^\mu )
	    =\partial_\mu\L^\mu \, .
$$
The variation of the action $\hat I$ with $\ve^-$ satisfying
the condition $\partial_-^3\ve^-=0$  yields
$$
\L^0={1\over\sqrt{2}}\biggl[ {\a\over 2}\partial_-\ve (\partial_-F
   -\partial_+F) + {\a^2\over 2}h\partial_-^2\ve + \ve\mu^2\biggr]\, .
$$
On the other hand, after using the equations of motion one obtains
$$
\D {\cal L}=\partial_\mu K^\mu\ \, ,\qquad
K^\mu\equiv {\partial {\cal L}\over \partial {\Phi^i}_{,\mu}}\d\Phi^i
- {\cal L}\ve^\mu \, ,
$$
so that the elimination of $F$ leads to
$$
K^0={1\over \sqrt{2}}\biggl\{ -\ve\bigl[ (1-h)T_{--}+{\a^2\over 2}
    \partial_-^2h-\mu^2\bigr] +{\a\over 2}\partial_-\ve (\partial_-F
    -\partial_+F +Q\partial_-h\bigr]\biggr\} \, .
$$
The Noether current takes the form
$$
{\cal N}^\mu = K^\mu -\L^\mu  \, ,
$$
which, after an explicite calculation, leads to
$$
{\cal N}^0 ={1\over \sqrt{2}}\biggl\{ \ve\bigl[ (h-1)T_{--}
  -{\a^2\over 2}\partial_-^2h\bigr] +{\a^2\over 2}\partial_
  -\ve\partial_-h -{\a^2\over 2}(\partial_-^2\ve )h\biggr\}\, .
							     \eqno(27)
$$
After decomposing $\ve$ as in Eq.(25a), ${\cal N}^0$ can be written
in the form
$$
{\cal N}^0=-{\a^2\over\sqrt{2}}(\o_-j^-+\o_0j^0 +\o_+j^+)\, ,\eqno(28a)
$$
where
$$\eqalign{
j^- &= (1-h)T_{--}+{1\over 2}\partial_-^2h  \, ,\cr
j^0 &= -{1\over 2}\partial_-h +\xi^-j^-  \, ,\cr
j^+ &= h+2\xi^-j^0-(\xi^-)^2j^-  \, .}                       \eqno(28b)
$$
The last equation yields
$$
h= j^+ -2\xi^-j^0 +(\xi^-)^2 j^-  \, .                       \eqno(29)
$$

These results will be helpfull in studying the Hamiltonian form of the
$SL(2,R)$ symmetry.
 The generators  $J^a$ of the global or residual symmetry of
the Hamiltonian equations of motion have to satisfy the following
conditions:

$-$ After the elimination of momenta they should be reduced to $j^a$,
i.e.
$$
J^a\bigl[ q,p(q) \bigr] = j^a(q,\dot q) \, .                \leqno(A)
$$

$-$ As discussed in details in the Appendix B, when the symmetry
parameters depend only of $\xi^+$, $\o_a =\o_a (\xi^+)$, then the
generators satisfy the relation
$$
\{ J^a,H_T\} +{\partial J^a\over\partial t}={\partial J^a\over
 \partial\s} \qquad {\hbox{i.e.}}\qquad \partial_-J^a=0 \, . \leqno(B)
$$

We shall now try to find the Hamiltonian currents $J^a$
{\it by using the Hamiltonian equations of motion analogous to}
(28b), with $j^a\to J^a$. These objects will authomaticaly satisfy
the condition (A).
Let us, therefore, define the Hamiltonian currents $J^a$ by
$$\eqalign{
J^- &= {\fr 1 2}\partial_-^2h           \, ,  \cr
J^0 &= -{\fr 1 2}\partial_-h +\xi^-J^-  \, ,  \cr
J^+ &= h + 2\xi^-J^0 - (\xi^-)^2J^-      \, ,  }             \eqno(30)
$$
where
$$
\partial_-X\equiv {1\over \sqrt{2}}\bigl[\{ X,H_T\} -X'\bigr] \, .
$$
An explicit calculation leads to the following result:
$$\eqalign{
J^-&= {1\over \a^2(h-1)}({\cal H}_0-{\cal H}_1)+{1\over \a^2}\mu^2 \cr
   &= {1\over\a^2}\biggl[ {2\over\a^2}(h-1)\pi_h^2-{2\over\a}F'\pi_h
       -{2\over\a}\pi_h\pi +2\pi_h'\biggr]  \, ,  \cr
J^0 &= \xi^-J^- +{\sqrt{2}\over\a^2}\biggl[ (h-1)\pi_h-{\a\over 2}F'
       -{\a\over 2}\pi \biggr]                   \, ,  \cr
J^+ &= h + 2\xi^-J^0 - (\xi^-)^2J^-  \, .     }              \eqno(31)
$$
Although these objects were obtained in the light-cone gauge, they can
be easily written in the gauge invariant form by making the transition
from $(h,\pi_h,F,\pi )$ to $(g_{11},\pi^{11},F,\pi_F)$. Their algebra
is gauge independent, but their dynamics is not.

If we now construct the generator $J$ in analogy with Eq.(28a), i.e.
$$
J =-{\a^2\over \sqrt{2}}\int d\s\, \bigl[
 \o_-(\xi^+)J^-  +\o_0(\xi^+) J^0 + \o_+(\xi^+)J^+ \bigr]\, ,\eqno(32a)
$$
we obtain the result
$$\eqalign{
J =\int d\s\, \biggl\{ &-{1\over\sqrt{2}}\ve\bigl[ {2\over\a^2}
   (h-1)\pi_h^2 -{2\over\a}F'\pi_h\bigr] + (\partial_+\ve
   -h\partial_-\ve )\pi_h \cr
 &+ ({\sqrt{2}\over\a}\ve\pi_h +{\a\over 2}\partial_-\ve )\pi
   +{\a\over 2}(\partial_-\ve )F' -{\a^2\over 2\sqrt{2}}
   (\partial_-^2\ve )h \biggr\} \, . }                       \eqno(32b)
$$
This quantity generates the transformations
$$
\d_0 h = \{ h,J\} \, ,\qquad \d_0 F = \{ F,J\} \, ,
$$
that coincide with (26) after the elimination of momenta.

The generator $J$ was found starting from the Noether current ${\cal
N}_0$ and replacing velocities with momenta using the Hamiltonian
equations of motion, in accordance with the condition $(A)$. Once we
have found $J$, we can derive the corresponding transformations of all
dynamical variables. Altough these transformations are the Hamitonian
analogue of the Noether symmetries of the action, it is instructive to
check whether they are the symmetries of the Hamiltonian equations of
motion. An analysis of Castellani's type (Appendix B) leads to the
consistency requirement $(B)$ on $J$. It can be easily verified that
the condition (B) is also fulfilled, so that $J$ {\it is indeed the
generator of the residual symmetry} of the Hamiltonian equations of
motion.

The importance of this result becomes evident when the currents $J^a$
are defined in a gauge independent way. In that case one can also find
the related $SL(2,R)$ transformations of the dynamical variables, but
the consistency condition $(B)$ tells us that these transformations are
the symmetries of the theory only in the light-cone gauge.

\subsection{Appendix: The conditions on the generators} 

The general method for studying the generators of local symmetries in
the Hamiltonian approach has been developed by Castellani. The method
yields an algorithm to construct the corresponding gauge generators.
A slight modification of the method can be applied to study the
generators of gauge symmetries in the case when the gauge parameters
are not arbitrary functions of space--time variables, but depend only
on $\xi^-$. In that case, there are no PFC constraints, so that the
equations of motion for $q^i$ and $p_i$ are

$$
{\dot q}^i=\{ q^i,H\} \, ,\qquad {\dot p}_i=\{p_i,H\} \, .
$$
If we demand that varied trajectories
$$
{\tilde q}^i\equiv q^i+\d_0 q^i \, ,\qquad
{\tilde p}_i\equiv p_i+\d_0 p_i     \, ,
$$
also satisfy the equations of motions, then
$$\eqalign{
\d_0{\dot q}^i &={\partial^2H\over \partial p_i\partial p_j}\d_0 p_j +
	{\partial^2H\over \partial p_i\partial q^j}\d_0 q^j \, ,   \cr
\d_0{\dot p}_i &= -{\partial^2H\over \partial q^i\partial p_j}\d_0 p_j-
	 {\partial^2H\over \partial q^i\partial q^j}\d_0 q^j \, . }
							     \eqno(A1)
$$
On the other hand, if the variations of $p$ and $q$ are produced by the
gauge generator,
$$
\d_0 q^i(\xi )=\{ q^i(\xi ), G \}   \, ,  \qquad
G\equiv\int d{\tilde\s}\, \o_a({\tilde\xi}^+) G^a({\tilde\xi})
							     \eqno(A2)
$$
[where ${\tilde\xi}=(\xi^0,{\tilde\s})$] it follows that
$$
\d_0{\dot q}^i=\int d{\tilde\s}\, \biggl[ \dot{\o}_a\{ q^i,G^a\}+
\o_a\{\{ q^i,G^a\},H_T\}+\o_a\{ q^i,{\partial G^a\over\partial t}\}
					       \biggr]\, ,   \eqno(A3)
$$
An analogous result holds for $\d{\dot p}_i$. Now, we can use the fact
that ${\dot\o}=\o'$ and perform the integration by parts in the first
term in (A3). If the surface term vanishes, we easily obtain the
relation
$$
\partial_-G^a\equiv\{ G^a,H\} +{\partial G^a\over\partial t}-(G^a)'=0
						   \, .      \eqno(A4)
$$

\subsection{References:}

\item{1.} A. M. Polyakov, Phys. Lett. {\bf B103} (1981) 207.
\item{2.} A. M. Polyakov, Mod. Phys. Lett. {\bf A2} (1987) 893;
V. G. Knizhnik, A. M. Polyakov and A. B. Zamolodchikov, Mod. Phys.
Lett. {\bf A3} (1988) 819.
\item{3.} N. D. Hari Dass and R. Sumitra, Int. J. Mod. Phys.
{\bf A4} (1988) 2245.
\item{4.}  F. David, Mod. Phys. Lett. {\bf A3} (1988) 1651;
J. Diestler and H. Kawai, Nucl. Phys. {\bf B321} (1989) 509.
\item{5.} Kai-Wen Xu and Chuan-Jie Zhu, Int. J. Mod. Phys. {\bf A6}
(1991) 2331;  Chang-Jun Ahn, Young-Jai Park, Kee Yong Kim and Yongduk
Kim, Phys. Rev {\bf D42} (1990) 1144; J. A. Helayel-Neto, S. Mokhtari
and A. W. Smith, Phys. Lett. {\bf B236} (1990) 12;  Chang-Jun Ahn,
Won-Tae Kim, Young-Jai Park, Kee Yong Kim and Yongduk Kim, Mod. Phys.
Lett. {\bf A7} (1992) 2263.
\item{6.} T. Kuramoto, Phys. Lett. {\bf B233} (1989) 363;
Y. Tanii, Int. J. Mod. Phys. {\bf 6A} (1991) 4639.
\item{7.} Ed. Sh. Egorian and R. P. Manvelian,  Mod. Phys. Lett.
{\bf A5} (1990) 2371.
\item{8.} E. Abdalla, M. C. B. Abdalla, J. Gamboa and A. Zadra,
Phys. Lett. {\bf B273} (1991) 222.
\item{9.} A. Mikovi\' c, Canonical quantization of 2d gravity coupled
to $c<1$ matter, Queen Mary and Westfield College preprint QMW/PH/91/22
(1992); Canonical quantization approach to 2d gravity coupled to $c<1$
matter, Imperial College preprint Imperial-TP/92-93/15 (1993).
\item{10.} S. Ghosh and S. Mukherjee, $SL(2,R)$ currents in 2D-gravity
are generators of improper gauge transformations, Saha Institute
preprint (1993).
\item{11.} P. A. M. Dirac, {\it Lectures on quantum mechanics\/,}
Yeshiva University --- Belfer Graduate School of Science (Academic,
New York, 1964); A. Hanson, T. Regge and C. Teitelboim,
{\it Constrained Hamiltonian Dynamics\/,} (Academia Nationale del
Lincei, Rome, 1976);
K. Sundermeyer, {\it Constrained Dynamics\/,} (Springer, Berlin, 1982).
\item{12.} J. Barcelos-Neto, Constraints and hidden symmetry in
2D-gravity, Univ. Rio de Janeiro preprint IF/UFRJ/92/21 (1992).
\item{13.} L. Castellani, Ann. Phys. (N. Y.) {\bf 143} (1982) 357.
\item{14.} R. Marnelius, Nucl. Phys. {\bf B211} (1983) 14.
S. Hwang, Phys. Rev. {\bf D28} (1983) 2614.
\item{15.} P. A. M. Dirac, Rev. Mod. Phys. {\bf 21} (1948) 392.
\item{16.}K. G. Wilson, Nucl. Phys. {\bf B} (Proc. Suppl.) {\bf 17}
(1990);  Prem. P. Srivastava, Constraints and Hamiltonian in
Light-Front Quantized Field Theory, preprint DFPF/9/TH/58 (1992).
\item{17.} T. Regge and C. Teitelboim, Ann. of Phys. (NY) {\bf 88}
(1974) 286; R. Benguria, P. Cordero and C. Teitelboim, Nucl. Phys.
{\bf B122} (1977) 61; P. J. Steinhardt, Ann. of Phys. (NY) {\bf 128}
(1980) 425.

\bye